\documentclass[journal]{IEEEtran}

\usepackage{booktabs} % For better looking tables
\usepackage{subcaption}
\usepackage{caption}
\usepackage{graphicx}
\usepackage{multirow}
\usepackage{comment}
\usepackage{amsmath}
\usepackage{hyperref}
\usepackage{url}
\def\UrlBreaks{\do\/\do-}
\expandafter\def\expandafter\UrlBreaks\expandafter{\UrlBreaks\do\a\do\b\do\c\do\d\do\e\do\f\do\g\do\h\do\i\do\j\do\k\do\l\do\m\do\n\do\o\do\p\do\q\do\r\do\s\do\t\do\u\do\v\do\w\do\x\do\y\do\z\do\A\do\B\do\C\do\D\do\E\do\F\do\G\do\H\do\I\do\J\do\K\do\L\do\M\do\N\do\O\do\P\do\Q\do\R\do\S\do\T\do\U\do\V\do\W\do\X\do\Y\do\Z\do\1\do\2\do\3\do\4\do\5\do\6\do\7\do\8\do\9\do\0}

\usepackage[svgnames]{xcolor}
\usepackage{url}

\begin{document}

\title{Subjective and Objective Analysis of Indian Social Media Video Quality}

\author{Sandeep Mishra,
        Mukul Jha,
        Alan C. Bovik,~\IEEEmembership{Fellow,~IEEE}% <-this % stops a space
\thanks{This work was supported by Mohalla Tech Pvt Ltd. A.C. Bovik was supported in part by the National Science Foundation AI Institute for Foundations of Machine Learning (IFML) under Grant 2019844. (Corresponding author: Sandeep Mishra.)}
\thanks{This work involved human subjects or animals in its research. Approval of all ethical and experimental procedures and protocols was granted by the Institutional Review Board (IRB), University of Texas, Austin, under FWA No. 00002030 and Protocol No. 2007-11-0066.}
\thanks{Sandeep Mishra, Alan C Bovik are with the
Department of Electrical and Computer Engineering, The University of Texas at Austin, TX 78712 USA (e-mail: sandy.mishra@utexas.edu, bovik@ece.utexas.edu). Mukul Jha is with ShareChat Inc., India
(e-mail: mukul@sharechat.co).}% <-this % stops a space
}

% The paper headers
\markboth{}%
{Mishra \MakeLowercase{\textit{et al.}}: Subjective and Objective Analysis of Indian Social Media Video Quality}

\maketitle

\begin{abstract}
We conducted a large-scale subjective study of the perceptual quality of User-Generated Mobile Video Content on a set of mobile-originated videos obtained from the Indian social media platform ShareChat. The content viewed by volunteer human subjects under controlled laboratory conditions has the benefit of culturally diversifying the existing corpus of User-Generated Content (UGC) video quality datasets. There is a great need for large and diverse UGC-VQA datasets, given the explosive global growth of the visual internet and social media platforms. This is particularly true in regard to videos obtained by smartphones, especially in rapidly emerging economies like India. ShareChat provides a safe and cultural community oriented space for users to generate and share content in their preferred Indian languages and dialects. Our subjective quality study, which is based on this data, offers a boost of cultural, visual, and language diversification to the video quality research community. We expect that this new data resource will also allow for the development of systems that can predict the perceived visual quality of Indian social media videos, to control scaling and compression protocols for streaming, provide better user recommendations, and guide content analysis and processing. We demonstrate the value of the new data resource by conducting a study of leading blind video quality models on it, including a new model, called MoEVA, which deploys a mixture of experts to predict video quality. Both the new LIVE-ShareChat dataset and sample source code for MoEVA are being made freely available to the research community at \href{https://github.com/sandeep-sm/LIVE-SC}{https://github.com/sandeep-sm/LIVE-SC}.  

\end{abstract}


\IEEEpeerreviewmaketitle

\section{Introduction}

\IEEEPARstart{A}{ccording} to a recent report by ETGovernment\cite{ETReport}, India is expected to surpass 1 billion internet users by 2030. Another report by Simon Kemp~\cite{Digital2023} states that approximately 33\% of the Indian population was active on social media as of January 2023. Indeed, a massive influx of new users has been observed on Indian social media platforms like ShareChat. According to statistics shared on their website~\cite{tierdata}, about 88\% of ShareChat users are from smaller, fast-growing (tier-2 and tier-3) cities. ShareChat provides a platform for users to generate content in their preferred languages and share them in a culturally oriented space that caters to a diverse population speaking more than 30 languages. The ensuing significant increases in social interactions entails the need to regulate and control their quality, especially in regards to the delivery of pictures and videos to the vast and diverse population of India. 

The report~\cite{Digital2023} also states that an overwhelming 97\% of these users prefer to browse social media platforms via their mobile phones, while the remaining 3\% use their laptops/desktops. Of course smartphones 1) are immediately available, 2) capture, record, and play audio and video, and 3) have wireless internet connectivity to download, upload and share data. Although smart mobile devices enable people to create and share their own videos, the perceptual quality of these videos depends heavily on their photographic skills and on the hardware and software specifications of their devices. Using low-end phones having poor camera capabilities can cause severe capture distortions, more so than high-end devices, but even these are subject to their users' uncertain skills, even with sophisticated software post-processing. All these factors introduce perceptual quality issues, which are reasonably represented in existing UGC-VQA databases, like LIVE-VQC~\cite{sinno2018large}, LSVQ~\cite{ying2021patch}, KoNViD~\cite{hosu2017konstanz}, and YouTube UGC~\cite{wang2019youtube}. 

While these data resources have significantly enabled UGC-VQA research, they are not very representative of the mixes and ranges of capture devices in India, where cheaper and lower-quality smartphone cameras are more prevalent than high-end smartphones. Moreover, contrary to the popular stereotype of India being a place only of vibrant colors, attires, weddings, and music, the video content that ordinary people post on social media platforms like ShareChat tends to be quite personal and evocative. There are many close-up shots of ordinary faces set against simple, plain backgrounds, with little attention paid to their attire and surroundings. The videos were often shot under challenging conditions, emphasizing the genuine, unadorned, and often haunting nature of the content.

Towards broadening the content and distortion diversity of the current suite of perceptual UGC-VQA datasets, we offer a new data resource called the LIVE-ShareChat IUGC-VQA Database, which is composed of $600$ mobile-originated UGC videos labeled with human perceptual quality judgments (Mean Opinion Scores, or MOS), and capturing a wide range of Indian cultural content and associated perceptual quality issues. Note that while we use the acronym UGC to refer to all user-generated content, when relevant we use the term IUGC to refer specifically to content captured by Indian users. We used this unique collection of video data to conduct a psychometric study under controlled laboratory conditions, where 48 subjects viewed 600 videos on a popular mobile device and rated their perceptual quality. A detailed synopsis of the study including its protocol and subsequent data analysis is given in Section \ref{sectionStudy}. As a way of demonstrating the value of this new psychometric data resource, we  evaluate and compare the performances of multiple legacy and state-of-the-art video quality prediction models on it in Section \ref{sectionBenchMark}. Further, we also introduce a new VQA model called \textbf{M}ixture-\textbf{o}f-\textbf{E}xperts \textbf{V}ideo-quality \textbf{A}ssessor (\textbf{MoEVA}), which significantly outperforms prior art models. Our contributions include:

\begin{itemize}
    \item Releasing the unique and diverse LIVE-ShareChat IUGC-VQA psychometric video quality database, which we envision will further empower VQA researchers on the development of practical UGC-VQA models able to perform well on wider varieties of user-generated content. Further, to the best of our knowledge, this is the first dataset built using only mobile-originated videos that were captured, edited, uploaded, and consumed on mobile devices. The human video quality ratings were also collected on a popular mobile device towards ensuring a viewing environment copacetic with prevalent real-world experiences.
    \item We conducted an extensive study and performance evaluation of existing VQA algorithms on the new LIVE-ShareChat IUGC-VQA database, demonstrating its practical research value while providing insights into the efficacies of existing video quality predictors. 
    \item We introduce a new mixture-of-experts based UGC-VQA model called MoEVA that combines spatial and temporal-statistical video analytics with content and quality aware features generated using a specialized pre-trained backbone. The effectiveness of MoEVA is demonstrated on the LIVE-ShareChat IUGC-VQA database.
    \item Both the dataset and new algorithm are being made freely available to VQA researchers at [insert Github/LIVE Link]
\end{itemize}

\begin{figure}[htbp]
    \centering
    {{\includegraphics[width=\linewidth]{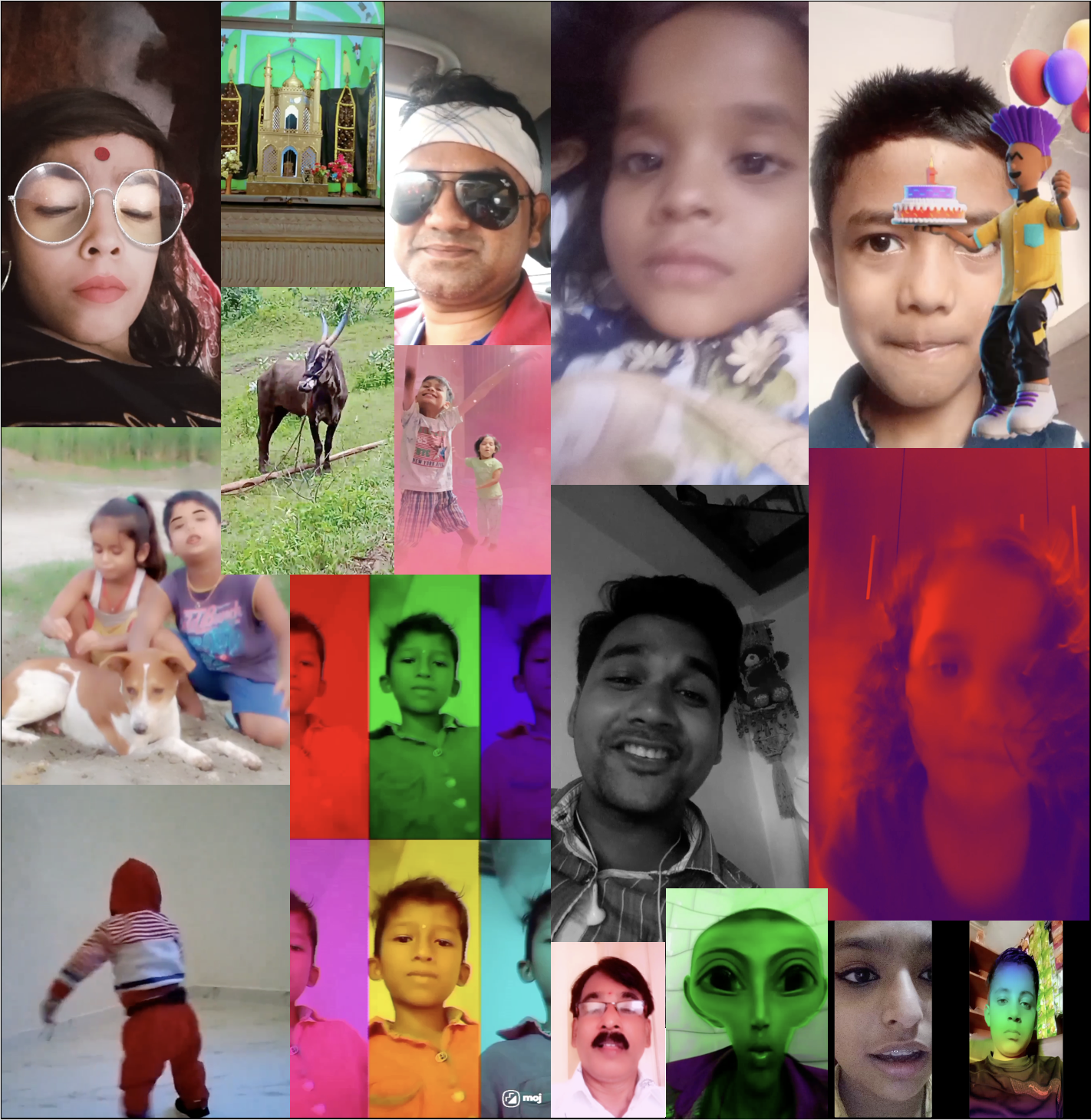}}}
    \captionsetup{justification=justified}
    \caption{{Exemplar frames from videos in the LIVE-ShareChat IUGC-VQA Database. There are many selfie-videos captured under low-light conditions, and many which applied various filters.}}
    \label{fig:ShareChat_collage}
\end{figure}

\section{Related Work}
\subsection{UGC-VQA Databases}
The Camera Video Database (CVD2014)~\cite{nuutinen2016cvd2014} was one of the first VQA databases relevant to the UGC-VQA scenario. It contains 234 videos, many captured with mobile devices, and many using DSLR cameras. Unfortunately, the videos in CVD2014 were post-processed by resizing to a fixed size, thereby introducing additional artifacts and hence the videos may no longer be regarded as real world content. The first true UGC VQA database, called the LIVE-Qualcomm Mobile In-Capture Database~\cite{ghadiyaram2017capture}, comprises 208 videos captured with a small number of smartphone capture devices. Neither of these datasets include very diverse content. A very large true UGC picture quality resource was soon introduced, called the LIVE In-the-Wild Image Quality Challenge Database~\cite{ghadiyaram2017live}, which crowdsourced 8000 human subjects who rated more than 1000 highly diverse UGC pictures. Inspired by this work, the authors of~\cite{hosu2017konstanz} created the KoNViD-1k VQA database, which contains 1200 UGC videos drawn from the YFCC100M dataset~\cite{thomee2016yfcc100m}. The videos were quality rated by 642 crowd workers. Unfortunately, these authors also resized the videos to a fixed spatial size destroying interpretation of the data as true UGC. A large true UGC VQA database called LIVE-VQC~\cite{sinno2018large}, comprises 585 videos scored by 4776 unique Amazon Mechanical Turk participants. The video contents in LIVE-VQC database were captured by 80 different photographers using their own mobile devices. The collected videos are true UGC without any post-capture processing. The YouTube-UGC Dataset~\cite{wang2019youtube} contains 1380 video clips rated by more than 8000 human subjects. The contents in the YouTube-UGC dataset were curated from videos uploaded on YouTube, and hence were meant for consumption on both computer/laptop screens and mobile devices. The largest and most comprehensive true UGC VQA database is the Large-Scale Social Video Quality Database (LSVQ)~\cite{ying2021patch} which contains about 39,000 videos, rated by more than 6000 unique subjects. LSVQ also contains human opinions collected on  117,000 space-time video patches and clips cropped from the original set of videos, making it useful for training deep learning models to learn space-time maps of video quality. Overall, LSVQ includes about 5M human judgments of UGC video quality.

\subsection{{UGC-VQA Models}}
 There are two broad categories of NR/Blind VQA (BVQA) algorithms: 1) Traditional VQA algorithms, which are feature-based, and which usually model certain statistical aspects of videos deemed predictive of MOS, and 2) Deep VQA models, trained on large datasets of human subjective judgments of video quality. There are also hybrid models, which use both statistical features and deep features.

\subsubsection{Traditional Models} Many older VQA models were distortion specific, and involved quantitative modeling of single distortions such as blockiness~\cite{wang2000blind}, blur~\cite{marziliano2002no}, ringing~\cite{feng2006measurement}, banding~\cite{wang2016perceptual}~\cite{tu2020bband}, and noise~\cite{amer2005fast}. More recent successful models involve mapping MOS of variably-distorted videos to general quality-aware features, using simple regressors such as support vector regressors (SVRs). Most popular traditional BVQA models deploy perceptually relevant low-level features derived from highly regular parametric bandpass neurostatistical video models~\cite{ruderman1994statistics}. The features used in these models are highly sensitive to arbitrary distortions and also are closely correlated with human visual distortion perception, making them particularly useful for VQA. For example, the popular NIQE~\cite{mittal2012making} model compares extracted Natural Video Statistics (NVS) features from an analyzed image to a set of "gold" NSS features (pre-computed on a set of pristine images), using a Mahalanobis distance measure to form quality estimates. BRISQUE~\cite{mittal2012no} uses the same NVS features to train a simple regressor (an SVR) to make video quality predictions. VIDEVAL~\cite{tu2021ugc} selectively fuses NVS features extracted from a variety of bandpass video models. 

\subsubsection{Deep Learning-based Models} Researchers have developed a variety of strategies to tackle the VQA problem using deep learning. Most of these use pre-trained backbones to extract image/video features, which are then fed to a fine-tuned regressor that makes video quality predictions. These models can be conveniently divided into two sub-categories based on the type of pre-training: 1) pre-training on an unrelated task such as object classification or segmentation, and 2) pre-training on a specific image/video quality assessment task. 

Among the first type is RAPIQUE~\cite{tu2021rapique}, a hybrid model that combines deep learning features extracted from a ResNet-50  backbone~\cite{he2016deep}, pre-trained on the ImageNet ~\cite{deng2009imagenet} classification task. The deep features computed on each frame supply quality-aware semantic information, which are pooled before feeding them into a simple regressor, along with a large set of perceptually relevant neurostatistical bandpass features. Bosse et al.~\cite{bosse2016deep} uses a deep network pre-trained on the ILSVRC~\cite{russakovsky2015imagenet} image classification dataset, then fine-tunes it to learn to predict image quality at the patch level. MUSIQ~\cite{ke2021musiq} uses a Vision Transformer (ViT) encoder pre-trained on the ImageNet classification task, fine-tuning it for quality prediction using an additional MLP layer.  

While the above models use deep features learned on an unrelated task to predict image/video quality, other models such as NIMA~\cite{talebi2018nima}, PQR~\cite{zeng2018blind}, MEON~\cite{ma2017end}, Patch-VQ~\cite{ying2021patch}, and CONTRIQUE~\cite{madhusudana2022image} are specifically trained on picture quality aware tasks. NIMA and PQR use raw human opinion scores to directly estimate the probability distributions of image quality scores, instead of using MOS. MEON employs a bifurcated architecture to handle two distinct yet interrelated tasks. The main task is predicting video quality, while the secondary task, which serves as extra supervision, is identifying the distortion.  Patch-VQ combines 2D spatial features extracted using the deep blind IQA model PaQ-2-PiQ~\cite{ying2020patches}, with 3D spatio-temporal features extracted using a 3D ResNet-18 pre-trained on the Kinetics dataset~\cite{kay2017kinetics}. By incorporating region pooling, Patch-VQ is able to produce local space-time maps of video quality. CONVIQT~\cite{madhusudana2022conviqt} is an unsupervised end-to-end deep learning-based model, which trains a ResNet-50 backbone using contrastive learning strategies to generate quality-aware features, which are mapped to MOS using a regularized ridge regressor.  As a demonstration of the utility of the LIVE-ShareChat IUGC-VQA database, we study the performance of these VQA models on it in Section \ref{sec:performanceofpriorart}.

\begin{figure}[htbp]
    \centering
    {{\includegraphics[width=\linewidth]{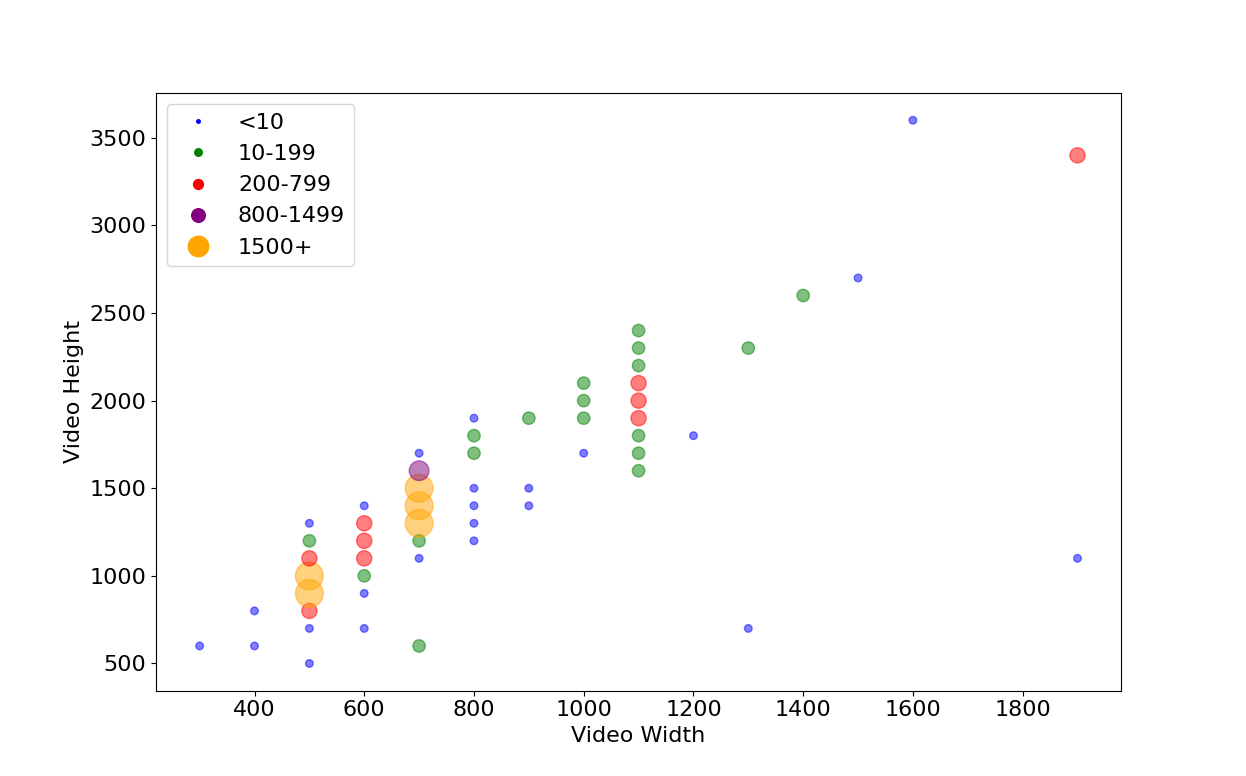}}}
    \captionsetup{justification=justified}
    \caption{Scatter plot of video widths versus video heights in the LIVE-ShareChat IUGC Video Quality dataset. The bubble sizes indicate the number of videos of each given dimension.}
    \label{fig:HxW_scatter}
\end{figure}

\begin{figure*}[htbp]
    \centering
    \begin{subfigure}{1\textwidth}
        \centering
        \includegraphics[width=\linewidth]{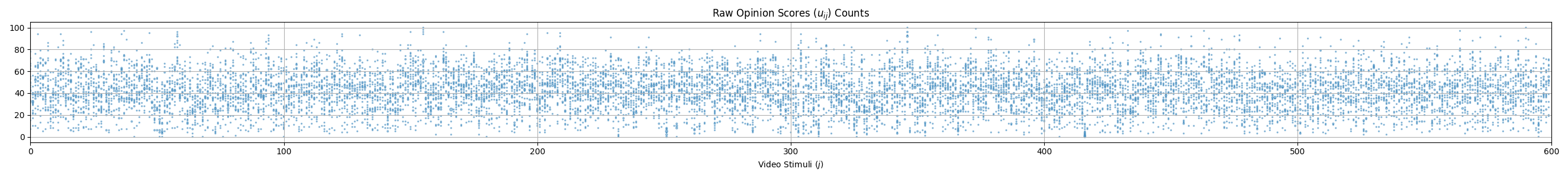}
        \caption{}
        \label{fig:rawVsMOS_sub1}
    \end{subfigure}%
    \vfill \hspace{0.1in}
    \begin{subfigure}{1\textwidth}
        \centering
        \includegraphics[width=\linewidth]{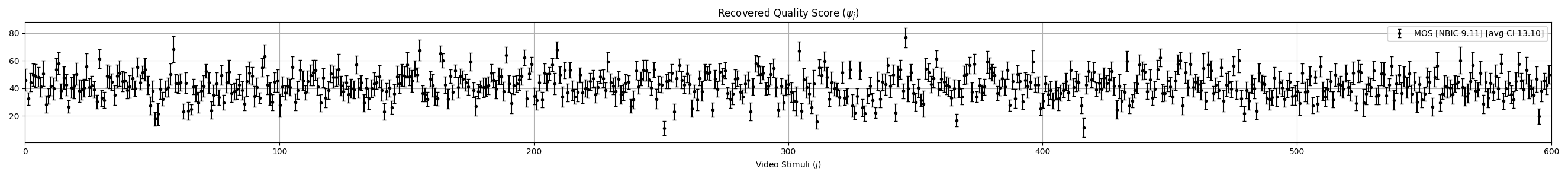}
        \caption{}
        \label{fig:rawVsMOS_sub2}
    \end{subfigure}
    \captionsetup{justification=justified}
    \caption{(a) Raw opinion scores vs (b) scores recovered using SUREAL.}
    \label{fig:rawVsMOS}
\end{figure*}

\section{Details of Subjective Study}
\label{sectionStudy}
\subsection{LIVE-ShareChat UGC-VQA Database}
The LIVE-ShareChat database contains 600 videos sampled from a publicly available set of 20,000 videos on the ShareChat website. The videos were pre-labeled by ShareChat video quality engineers with annotations pertaining to quality issues commonly found in user-generated content, such as jitter and blur, abnormal lighting, excess camera movement, etc. We ensured that each type of annotated issue was well represented. The dimensions of each video depend on the camera specifications, the settings during capture, and any editing by the user. The heights of the original superset of videos varied between 528 to 5428 pixels, while the widths varied between 320 to 2420 pixels. To make our ultimate data more amenable for training and processing by existing learning architectures, among these we only used videos having heights between 900 to 1500, and widths between 500 to 800 pixels. As shown in Figure \ref{fig:HxW_scatter}, the frequency distributions of the size dimensions lie within these ranges. All of the videos have heights greater than widths, making them suitable for viewing in portrait mode, which is preferred on social media platforms like ShareChat, Instagram, TikTok, etc. The videos were selected to have durations lying between 10 and 65 seconds, then were clipped to 8 seconds, in accordance with ITU-T P.913 Section 6.5~\cite{ITUT}. This choice allowed us to present more different contents to each user while reducing temporal quality variations, making it easier for the subjects to provide overall video quality judgments. 

\subsection{Subjective Study Environment}
The human study was conducted in the Subjective Study room at the Laboratory for Image and Video Engineering (LIVE) at The University of Texas at Austin. Since the majority of social media users browse social UGC videos on mobile devices, we used a Google Pixel 5 with the Android 11 operating system to display the videos, using a special purpose Android application. The device has a 6" inch OLED panel with FHD+ resolution supporting a refresh rate of up to 90Hz. We fixed the brightness of the device to 75\% of the maximum to avoid the occurrence of automatic changes during the study. The device also supports automatic re-scaling of videos to fit the screen, thus removing any requirement on our end to do so. While this is a processing step, it is an inevitable one, and so it reflects the content that would be viewed by people during normal use. Hence the videos were displayed to subjects just as they would have viewed them on their own devices. We connected an external keyboard and a mouse to facilitate stable viewing and to simplify the rating process.

The study room is both sound and light-proof to mimic an isolated environment. We ensured that the artificial lighting arrangements did not interfere with the viewing conditions by placing them at strategic locations to simulate a living room lighting environment. The incident luminance on the mobile screen was measured to be approximately 150 lux. The device was securely stationed on a smartphone mount with adjustable viewing angles, and a height-adjustable chair was provided to the subjects to comfortably position themselves. The subjects were asked to sit at a distance of about three-fourths of their arm's length to mimic typical social media browsing behavior. They were also asked to avoid making significant changes to their seating and viewing arrangements once the study began. We directed them not to alter any device settings, and to use the keyboard and mouse to communicate with the application only when prompted. 

Upon arrival, each participant was assigned a subject number as an identifier, and a predefined playlist of videos was played for them. After each video playback, a rating screen appeared with a rating bar for the subject to provide their evaluation of subjective video quality. The rating bar represents a continuous 0-100 scale, based on the SAMVIQ scale suggested in ITU-T P.913 Section 7.1.4. Five Likert labels marked the bar: Bad(0), Poor(25), Fair(50), Good(75), and Excellent(100), where the score in ($\cdot$) are numerical scores associated with the label but not visible to the subjects. The initial position of the cursor was set to 0 and the subject was guided to use the wireless mouse to move the cursor to their desired score. Once each subject finalized a score, they were directed to press the NEXT button, whereby their score was recorded in a text file while the next video playback began. The application did not allow replaying videos since we wanted to record only the instinctual responses of the participants. Lastly, the subjects were guided to avoid any distractions throughout the study. 

\begin{figure}[htbp]
    \centering
    {{\includegraphics[width=\linewidth]{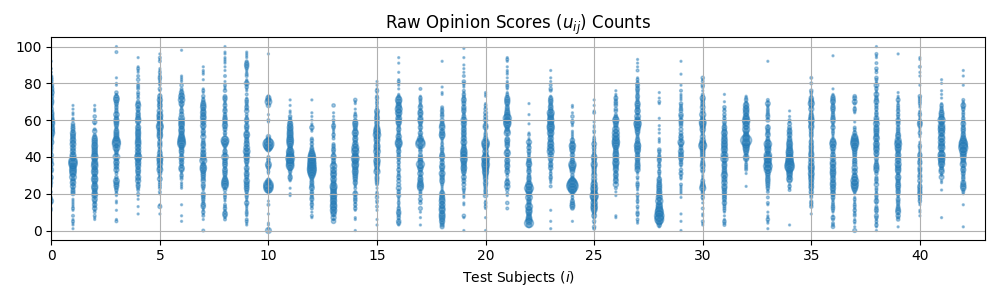}}}
    \captionsetup{justification=justified}
    \caption{Distributions of raw opinion scores.}
    \label{fig:raw_opinion_scores}
\end{figure}

\begin{figure*}[htbp]
    \centering
    \begin{subfigure}{.301\textwidth}
        \centering
        \includegraphics[width=\linewidth]{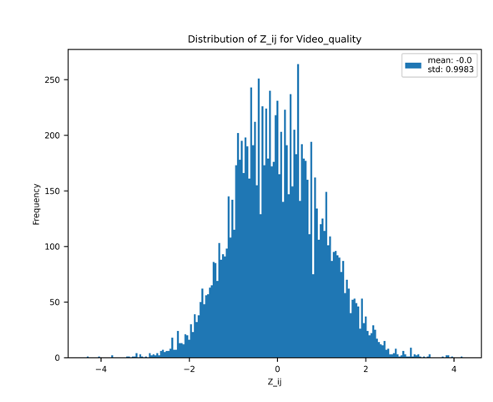}
        \caption{}
        \label{fig:MOS_dist_sub1}
    \end{subfigure}%
    \hfill \hspace{0.1in}
    \begin{subfigure}{.327\textwidth}
        \centering
        \includegraphics[width=\linewidth]{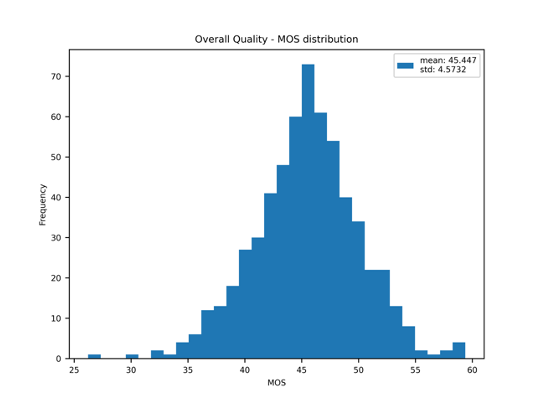}
        \caption{}
        \label{fig:MOS_dist_sub2}
    \end{subfigure}%
    \hfill
    \begin{subfigure}{.327\textwidth}
        \centering
        \includegraphics[width=\linewidth]{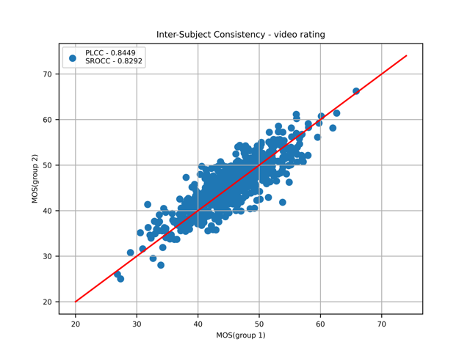}
        \caption{}
        \label{fig:MOS_dist_sub3}
    \end{subfigure}
    \captionsetup{justification=justified}
    \caption{(a) Histogram of z-scores of human subjective judgments. (b) Histogram of MOS. (c) scatter plot of inter-subject consistency trials.}
    \label{fig:MOS_dist}
\end{figure*}

\subsection{Subjective Testing Protocol}

We followed a single stimulus testing protocol as described in ITU-T P.913 Section 3.2.13. Since the LIVE-ShareChat dataset contains user-generated content, it does not involve the concept of reference and distorted videos. The dataset contains a total of 600 videos with a viewing time of about 8 seconds per video, resulting in a total of 4,800 seconds of playback time. We estimated that subjects would require about 20 seconds to view and rate each stimulus, for a total of 12,000 seconds or 3.33 hours over the entire dataset. To reduce the time required of each volunteer, we divided the dataset into four non-overlapping playlists, each containing 150 videos. The 48 subjects were evenly divided into 4 groups of 12 each. Each group was assigned two of the four playlists in a round-robin fashion. As a result, each volunteer viewed two playlists of 150 videos each, resulting in a total per-subject viewing time of 6,000 seconds. We split this into two sessions per subject and played a single playlist in each session. Each subject was required to attend two sessions to complete the study. We ensured there was a minimum gap of 24 hours between each subject's two sessions to reduce fatigue. Given that each playlist was viewed by 24 subjects, each video in our dataset was labeled with 24 ratings.

\subsection{Subject Screening and Training}
We recruited 48 volunteers having varying academic backgrounds from the student community of The University of Texas at Austin. The volunteer pool generally had little/no experience in video quality evaluation. Each subject participated in 2 sessions conducted on different days.  

Visual acuity and color perception tests were conducted on each volunteer to identify any deficiencies. We conducted the Ishihara Color blindness test and found one color-blind participant. The Snellen's Eye test determined that every subject had 20/20 vision while wearing their corrective glasses/contact lenses, if any. We did not exclude any volunteers based on these outcomes. 

Subjects were then introduced to the study room, the setup, the purpose of the study, and the nature of the videos they would be viewing. We instructed them to only rate perceived visual quality while ignoring the nature of the contents. To familiarize each subject with the type of videos they would be viewing, before the beginning of each session they participated in a short training session, where they were shown three different videos of different contents and diverse qualities, which they were prompted to rate as practice. The scores recorded during the training session were not included in the psychometric database. 

\subsection{Post Study Questionnaire}
At the conclusion of each rating session, each subject was asked to fill out a questionnaire. This data was collected to record demographics, feedback on the study protocol, and comments on the ease of participation.

Approximately 85\% of the subject population was male and the rest female. The minimum age of the subject pool was 21, while the maximum was 29. The mean, median, and standard deviation of the ages of the participants was found to be 24.75, 24.0, and 2.34. More than 90\% of the pool felt that 8 seconds of visual playback was enough to adequately judge the quality of the videos. None of the participants complained about any dizziness during their sessions. 

\begin{figure*}[htbp]
    \centering
    {{\includegraphics[width=\textwidth]{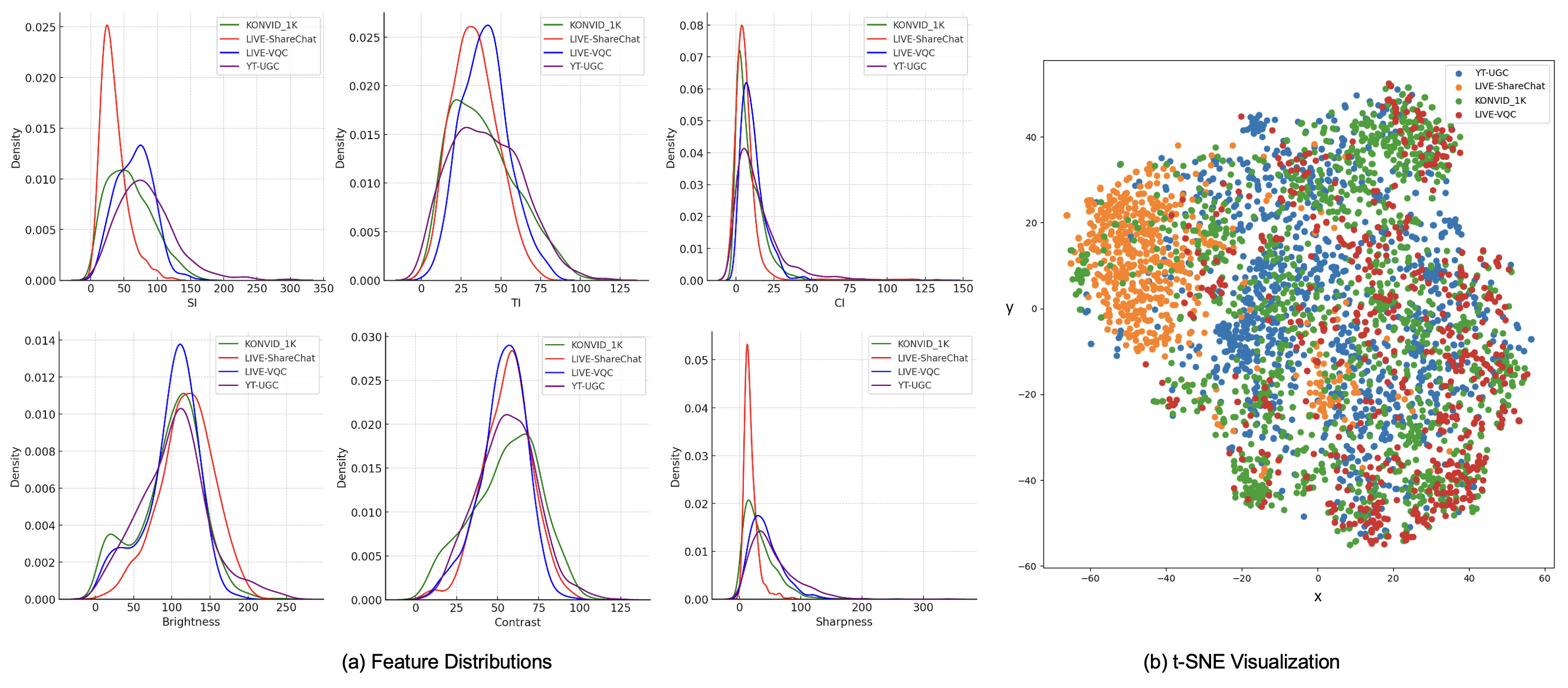}}}
    \captionsetup{justification=justified}
    \caption{(a) Comparison of the six feature distributions among the 4 considered datasets: KoNViD-1k, LIVE-ShareChat, LIVE-VQC, YouTube-UGC. (b) t-SNE visualization of the VGG19 features of the four datasets.}
    \label{fig:freq_dist_and_tSNE}
\end{figure*}

\subsection{Processing of Subjective Scores}
We studied the reliability of the recorded opinion scores by conducting an analysis of the inter-subject and intra-subject consistencies of the raw opinion scores collected during the study. \\

\subsubsection{Inter-subject consistency} To calculate the inter-subject consistency of the subject data, we divided the scores recorded on each video into two groups of equal size, then measured the correlation of MOS between the two groups. We repeated this process over 100 random divisions and computed the median PLCC (Pearson linear correlation coefficient), which was found to be 0.85, while the median SROCC (Spearman rank order correlation coefficient) was found to be 0.83. The results are tabulated in Table \ref{tab:ConsistencyScores}.\\

\subsubsection{Intra-subject consistency} To quantify the intra-subject consistency we compute the PLCC and SROCC between each individual's opinion scores and the MOS. The median PLCC was 0.62 and the median SROCC was 0.60. The results are also given in Table \ref{tab:ConsistencyScores}. Since the data is UGC these figures are not as high as commonly occurs on synthetically generated video distortions data \cite{sinno2018large, ebenezer2023hdr, saha2023study}, but are typical of other UGC datasets \cite{ying2020patches, ying2021patch}.

\begin{table}[htbp]
    \centering
    \begin{tabular}{lcc}
        \toprule
        & SRCC & PLCC \\
        \midrule
        Inter-Subject Consistency & 0.8292 & 0.8459 \\
        Intra-Subject Consistency & 0.5925 & 0.6154 \\
        \bottomrule
    \end{tabular}
    \caption{Consistency Scores}
    \label{tab:ConsistencyScores}
\end{table}

We also conducted a subject rejection protocol to eliminate unreliable subjects, using the effective SUREAL method described in~\cite{li2017recover}, which is less susceptible to subject corruption and provides tighter confidence intervals than prior methods.

SUREAL~\cite{li2017recover} models the raw opinion scores of videos as random variables $\{X_{e,s}\}$ having the following form:

\begin{align}
    X_{e,s} &= x_e + B_{e,s} + A_{e,s} \nonumber \\
    B_{e,s} &\sim  \mathcal{N} (b_s, v^2_s) \\
    A_{e,s} &\sim  \mathcal{N} (0, a^2_{c:c(e)=c}) \nonumber 
\end{align}

\noindent where $e = 1, 2, 3, ..., 600$ are the indices of the videos, $s = 1, 2, 3, ..., 48$ are the unique human participants, and $x_e$ is the quality of the video $e$ as perceived by a hypothetical unbiased and consistent viewer. $B_{e,s}$ are i.i.d Gaussian variables representing the human subject $s$, parameterized by a bias (mean) $b_s$ and inconsistency (variance) $v^2_s$, which are assumed constant across all videos viewed by the subject $s$. $A_{e,s}$ are i.i.d Gaussian variables representing a particular video content parameterized by the ambiguity (variance) $a^2_c$ of content $c$, and $c = 1, 2, ...600$ indexes the unique source sequences in the database. Content ambiguity is assumed constant for each video, independent of the subject viewing it, and since each video is unique, the $A_{e,s}$ are unique. In this formulation, the parameters $\theta = ({x_e}, {b_s}, {v_s}, {a_c})$ are the variables of the model. To estimate $\theta$, the log-likelihood function $L$ is defined :

\begin{equation}
    L = \log P(\{x_{e,s}\}|\theta).
\end{equation}

Using the data obtained from the psychometric study, we derived a solution $\hat{\theta} = $ argmax$_\theta L$ using the Belief Propagation algorithm~\cite{li2017recover}. 

Next, we describe how we obtained MOS. Let $m_{ijk}$ denote the score recorded for video $j$ provided by subject $i$ in session $k = {1, 2}$. Let $\delta(i, j)$ be the indicator function

\begin{equation}
    \delta(i, j) = \begin{cases} 
      1 & \textrm{if subject } i \textrm{ rated video } j, \\
      0 & \textrm{otherwise}
      \end{cases}
\end{equation}

\noindent which is required since not all videos in the database are rated by every subject. We calculated the normalized opinion scores received across multiple sessions of each subject as 

\begin{align}
    \mu_{ik} &= \frac{1}{N_{ik}}\sum^{N_{ik}}_{j=1}{m_{ijk}} \nonumber \\
    \sigma_{ik} &= \sqrt{\frac{1}{N_{ik}-1}\sum^{N_{ik}}_{j=1}{(m_{ijk}-\mu_{ik})^2}} \nonumber \\
    z_{ijk} &= \frac{m_{ijk}-\mu_{ik}}{\sigma_{ik}} \nonumber
\end{align}

\noindent where $z_{ijk}$ are the per-session normalized opinion scores (z-scores) and $N_{ik}$ is the number of videos seen by subject $i$ in session $k$. The z-scores over all sessions were concatenated to form the matrix $\{z_{ij}\}$ denoting the z-score assigned by subject $i$ to the videos indexed by $j$ with $j \in \{1, 2, \ldots, 600\}$, where the entries of $\{z_{ij}\}$ are empty at locations $(i, j)$ where $\delta(i, j) = 0$. Assuming the $z_{ij}$ to have a standard normal distribution, $99\%$ of the z-scores were found to lie in $[-5, 5]$. The scores were linearly mapped to the range $[0, 100]$:
\begin{equation}
    z^{'}_{ij} = \frac{100(z_{ij} + 5)}{10},
\end{equation}

 \noindent and finally, the Mean Opinion Score (MOS) of each video was calculated by averaging the scores received on that video:

\begin{equation}
    MOS_j = \frac{1}{N_{j}}\sum^{N}_{i=1}{z^{'}_{ij}\delta(i,j)},
\end{equation}

\noindent where $N_j = \sum^{N}_{i=1} \delta(i, j)$. The correlation between the scores obtained by SUREAL and by the traditional method (ITU~\cite{SubRej}) was 0.996.

The MOS was found to lie in the range [26.19, 65.66], and the mean standard deviation of rescaled z-scores over all subjects and all videos was found to be 6.99. The histogram of MOS is shown in Fig. \ref{fig:MOS_dist}, indicating a nicely regular distribution of MOS.

\begin{figure}[htbp]
    \centering
    {{\includegraphics[width=\columnwidth]{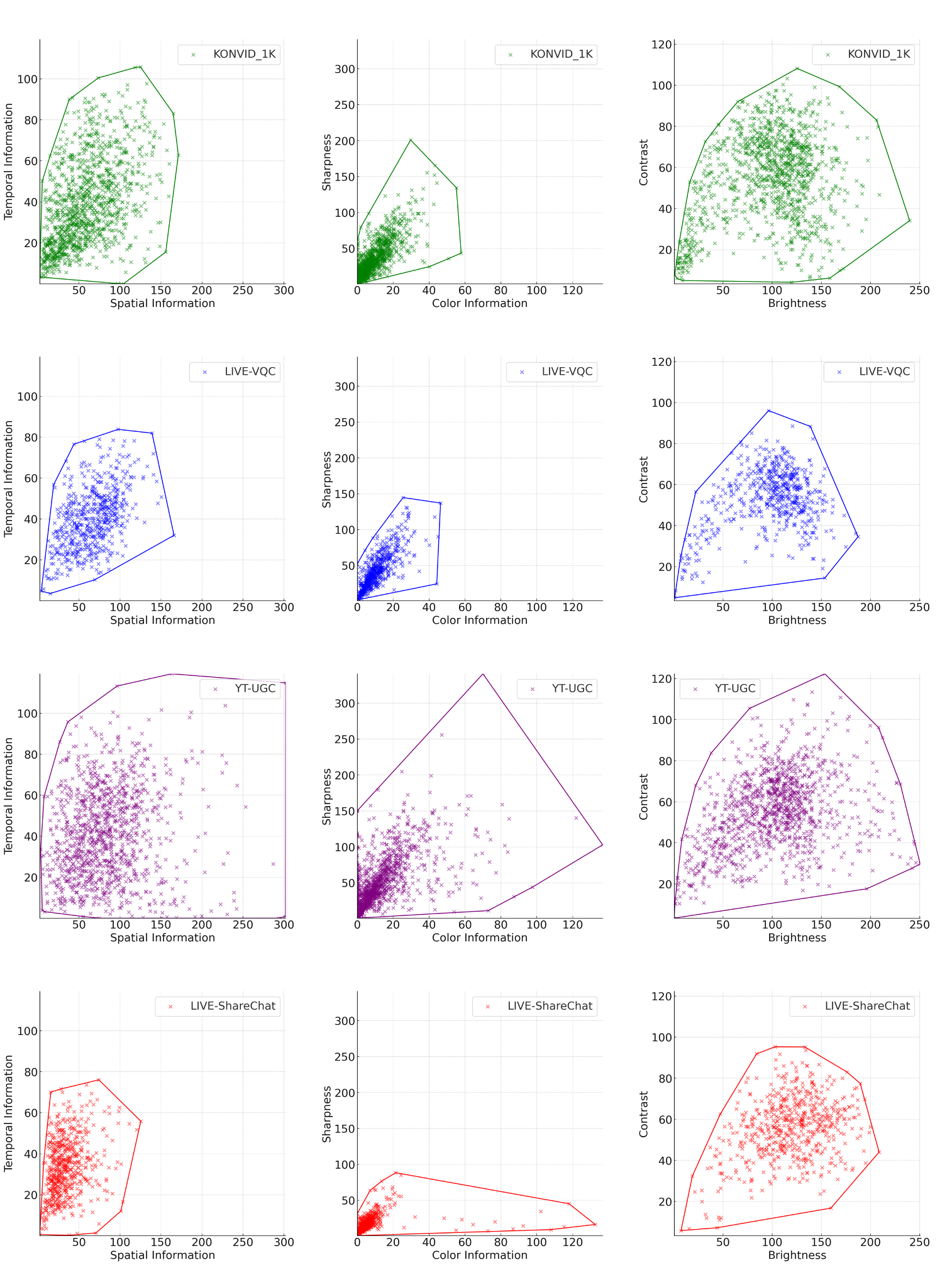}}}
    \captionsetup{justification=justified}
    \caption{Content distribution (\textcolor{Green}{green}: KoNViD-1k, \textcolor{blue}{blue}: LIVE-VQC, \textcolor{violet}{purple}: YouTube-UGC, \textcolor{red}{red}: LIVE-ShareChat) in paired feature space with their corresponding convex hulls. Column 1: SI x TI, Column 2: CI x Sharpness, Column 3: Brightness x Contrast.}
    \label{fig:diversity_plots}
\end{figure}

\subsection{{Analysis and Visualization of Opinion Scores}}
Fig. \ref{fig:rawVsMOS}(a) depicts the raw opinion scores collected on each video. Fig. \ref{fig:rawVsMOS}(b) is a box plot of the MOS recovered using SUREAL, where the endpoints of $MOS_j$ lie at a distance of $\pm STD_j$, the standard deviation of the normalized opinion scores used to recover $MOS_j$. Fig. \ref{fig:raw_opinion_scores} plots the per-subject distributions of human opinion scores, illustrating the need for normalization (z-scores). We also plotted the histogram of the scores $z_{ij}$ recovered using SUREAL, and the histogram of the recovered MOS in Figs. \ref{fig:MOS_dist}(a) and \ref{fig:MOS_dist}(b), respectively. The mean of the recovered opinion scores was found to be 45.45 and the standard deviation was found to be 4.57. The distribution is unimodal and well distributed. The fairly narrow MOS distribution suggests greater difficulty of segregating between different video quality levels, presenting challenges to prediction models. This is typical of UGC videos containing real-world distortions, in contradiction to datasets of videos that have been altered by very wide ranges of severity of synthetically applied distortions~\cite{ying2020patches}. Moreover, the complex interplay between UGC video contents and (typically) multiple coincident distortions makes the problem even more difficult.

\begin{figure}[htbp]
    \centering
    {{\includegraphics[width=0.8\columnwidth]{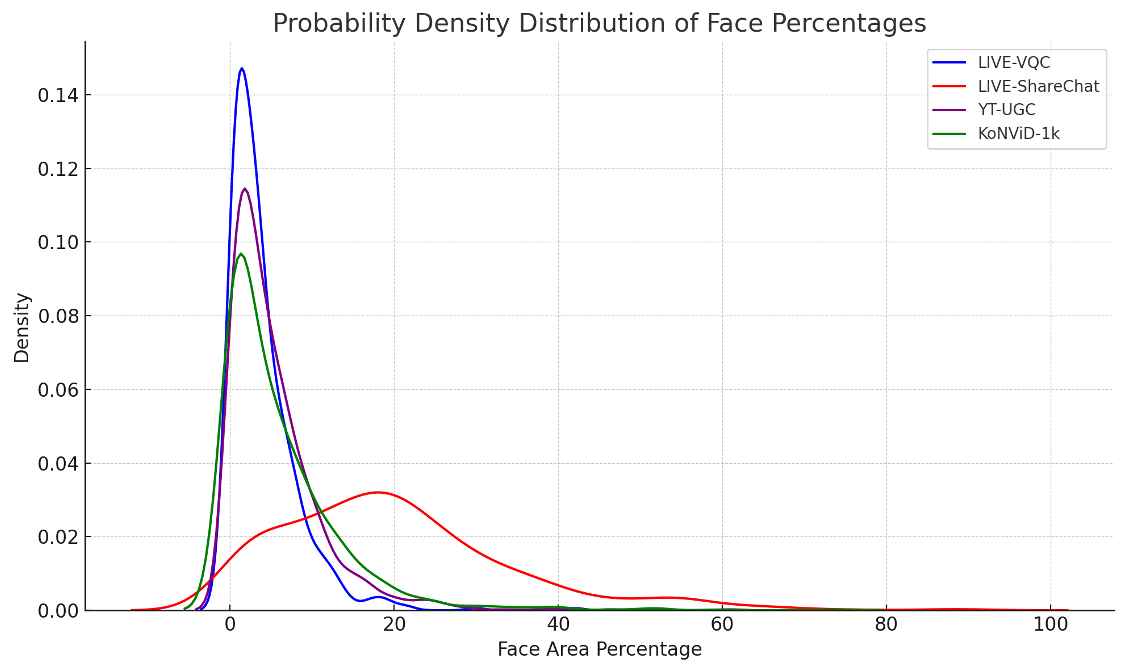}}}
    \captionsetup{justification=justified}
    \caption{Comparison of the distributions of Face Area Percentages of the four compared UGC-VQA datasets.}
    \label{fig:fap}
\end{figure}

\subsection{{Content Diversity and MOS Distribution}}
To study the degree of content diversity in the new dataset, and to compare with other UGC-VQA datasets, we measure six quantitative attributes related to spatial and temporal characteristics, as suggested in \cite{tu2021ugc}. These include brightness, contrast, colorfulness (CI), sharpness, spatial information (SI), and temporal information (TI) as described in Eq. \ref{eq:grouped}, where $V_f$ is the $f^{th}$ frame of the video $V$, $V_{fhw}$ is the luminance value of the pixel located at coordinate $(h,w)$ in the $f^{th}$ frame of the video, $rg$ is the difference between the red and green channels, $yb$ is the difference between the yellow and blue channels where the yellow channel is computed as the average of the red and green channels, and $\sigma, \mu$ are the sample standard deviation and mean operators. 
\begin{align}
    \text{Brightness} &= \frac{1}{H*W*F}\sum^{F}_{f=1}\sum^{H}_{h=1} \sum^{W}_{w=1} V_{fhw}\notag \\
    \text{Contrast} &= \frac{1}{F}\sum^{F}_{f=1} \sigma \left(V_f\right) \notag \\
    \text{Sharpness} &= \frac{1}{(H-2) \times (W-2) \times F} \notag \\
    &\times \sum_{f=1}^{F} \sum_{h=2}^{H-1} \sum_{w=2}^{W-1}\sqrt{ \left( \frac{\partial V_{fhw}}{\partial h} \right)^2 + \left( \frac{\partial V_{fhw}}{\partial w} \right)^2 } \notag \\
    \text{SI} &= \frac{1}{F} \sum_{f=1}^{F} \sigma \left( \sqrt{ \left( \frac{\partial V_{f}}{\partial x} \right)^2 + \left( \frac{\partial V_{f}}{\partial y} \right)^2 } \right) \label{eq:grouped} \\
    \text{TI} &= \frac{1}{F-1} \sum_{f=2}^{F} \sigma \left( V_{f} - V_{f-1} \right) \notag \\
    \text{CI} &= \frac{1}{F} \sum_{f=1}^{F} \left( \sqrt{{\sigma(rg_{f})}^2 + {\sigma(yb_{f})}^2} \right. \notag \\
    &\left. + 0.3 \sqrt{\mu(rg_{f})^2 + \mu(yb_{f})^2} \right) \notag \\
    \nonumber
\end{align}
The features were calculated on every $10^{th}$ frame, to limit computation, then averaged across frames for each content. We compared the features computed on the LIVE-ShareChat IUGC-VQA database with those computed on three large-scale UGC-VQA databases: LIVE-VQC, KoNViD-1K, and YouTube-UGC. Fig. \ref{fig:freq_dist_and_tSNE} (a) plots the distributions of each feature for each dataset. We also drew comparative feature space visualizations, along with their corresponding convex hulls, by pairing SI with TI, CI with sharpness, and brightness with contrast as shown in Fig. \ref{fig:diversity_plots}. We also extracted VGG-19 \cite{simonyan2014very} features on each video and visualized them in a 2D subspace using t-SNE \cite{maaten2008visualizing} (Fig. \ref{fig:freq_dist_and_tSNE} (b)).

It may be observed that the cultural videos in the LIVE-ShareChat IUGC-VQA database have divergent characteristics, there being a higher density of videos having lower SI, TI, CI, and Sharpness as compared to the videos in LIVE-VQC, KoNViD-1k, and especially YouTube-UGC. This difference is also observed on the t-SNE features, where the LIVE-ShareChat videos form a separate cluster having little overlap with the features from the other datasets. It is evident from Fig. \ref{fig:diversity_plots} that features derived from the LIVE-ShareChat videos exhibit the least coverage of SI-TI space, whereas Youtube-UGC shows the most coverage. These cultural videos have simpler content with less "action". The CI and Sharpness features computed on the ShareChat videos exhibit a skewed pattern, with most of the videos having lower values of both features. In contradiction to any possible expectations that Indian content would be highly colorful, they instead tend to be more subdued. However, the LIVE-ShareChat videos exhibit similar coverages of brightness and contrast as the other datasets.

The LIVE-ShareChat database contains many videos with face close-ups, quite unlike other datasets where the variation of depth of field and types of content is much higher. To validate this observation, we plotted the distribution of Face Area percentage of the videos in each dataset in Fig. \ref{fig:fap}. The Face Area percentage was computed on every $10^{th}$ frame using the state-of-the-art face detector MTCNN \cite{zhang2016joint}, then averaged over frames. 

\section{Benchmarking Objective NR-VQA Algorithms}
\label{sectionBenchMark}
\begin{table*}[htbp]
    \centering
    \begin{tabular}{lcccc}
        \toprule
        Method & SRCC $\uparrow$ & KRCC $\uparrow$ & PLCC $\uparrow$ & RMSE $\downarrow$ \\
        \midrule
        NIQE & 0.3954 & 0.2276 & 0.3288 & 5.0920 \\
        BRISQUE & 0.4766 & 0.3340 & 0.4922 & 4.8439 \\
        VIDEVAL & 0.7104 & 0.5172 & 0.7087 & 3.8681 \\
        PQR & 0.6930 & 0.5054 & 0.6836 & 3.9119 \\
        RAPIQUE & 0.7280 & 0.5410 & 0.7392 & 3.7194 \\
        Patch-VQ & 0.7120 & 0.5205 & 0.6999 & 3.7806 \\
        Li \textit{et al.} & 0.7189 & 0.5357 & 0.7212 & 3.7431 \\
        CONTRIQUE & 0.7154 & 0.5254 & 0.7241 & 3.8120 \\
        CONVIQT & 0.7210 & 0.5289 & 0.7299 & 3.7314 \\
        CONTRIQUE + S-NSS + T-NSS & 0.7353 & 0.5436 & 0.7403 & 3.7153 \\
        \midrule
        \textbf{MoEVA} & \textbf{0.7524} & \textbf{0.5626} & \textbf{0.7599} & \textbf{3.5932} \\
        \bottomrule
    \end{tabular}
    \caption{Performance evaluation of NR-VQA algorithms on the LIVE-ShareChat IUGC database.}
    \label{tab:performance_VQA}
\end{table*}

As a way of demonstrating the scientific usefulness of the new LIVE-ShareChat IUGC database, we used it to evaluate the efficacies of a number of publicly available No-Reference (NR-VQA) algorithms. We were also interested in the impact of the unique characteristics of the database on existing VQA algorithms. We selected six well-known general-purpose NR-VQA models to test: NIQE~\cite{mittal2012making}, BRISQUE~\cite{mittal2012no}, VIDEVAL~\cite{tu2021ugc}, RAPIQUE~\cite{tu2021rapique}, and CONTRIQUE~\cite{madhusudana2022image}. Among these, NIQE and BRISQUE were created as image quality assessment algorithms but have been widely deployed for VQA. We adapted them for videos by simply average-pooling over time the quality-aware features extracted individually from each frame. When adopting the training-free NIQE model, the predicted frame quality scores were pooled to yield the final video quality scores. For the methods that require training (BRISQUE, VIDEVAL, RAPIQUE, and CONTRIQUE), we trained a support vector regressor (SVR) with the radial basis function kernel to learn mappings from the pooled quality-aware features to the ground truth MLE-MOS. VIDEVAL was designed by carefully curating 60 statistical features having high correlation with human quality judgments. RAPIQUE combines natural scene statistics with quality-semantic deep learning features. CONTRIQUE uses contrastive pre-training to learn features associated with image distortion classification, that are then used to train a model without supervision to conduct quality assessment. We evaluated the performance of these objective NR-VQA algorithms using the following metrics: Spearman’s Rank Order Correlation Coefficient (SROCC), Kendall Rank Correlation Coefficient (KRCC), Pearson’s Linear Correlation Coefficient (PLCC), and Root Mean Square Error (RMSE). The metrics SROCC and KRCC measure the monotonicity of the objective model predictions against human scores, while the metrics PLCC and RMSE measure prediction accuracy. The predicted quality scores were passed through a logistic non-linearity [38] to further linearize the objective predictions and to place them on the same scale as MOS: 

\begin{equation}
    f(x) = \beta_2 + \frac{\beta_1 - \beta_2}{1 + \exp(-x + \beta_3/ |\beta_4|)}
\end{equation}

We tested the VQA algorithms mentioned above on 1000 random train-test splits using the four metrics. For each split, 80\% of the videos were randomly chosen for training and validation, while the remaining 20\% constituted the test set. All of the algorithms were tested on the test set after pre-training them on the training set generated using the aforementioned train-test split, except NIQE, which does not require any pre-training. Since NIQE is an unsupervised model, we evaluated its performance on all 1000 test sets, without any training. We applied five-fold cross-validation to the training and validation sets of BRISQUE, VIDEVAL, RAPIQUE, and CONTRIQUE to find the optimal parameters of the SVRs they were built on.

\subsection{Performance of NR-VQA Models}
\label{sec:performanceofpriorart}
Table \ref{tab:performance_VQA} lists the performances of the aforementioned NR-VQA algorithms on the LIVE-ShareChat UGC database. We found that NIQE performed poorly since it is not trained and since it was developed using a set of pristine images available at the time of its development. Over time, the quality and characteristics of cameras and camera processing pipelines have changed. However, the performance of BRISQUE was slightly superior. Note that BRISQUE uses the same features as NIQE to train an SVR head to regress quality scores. The performance of VIDEVAL, RAPIQUE, and CONTRIQUE was much better than NIQE or BRISQUE. In the case of VIDEVAL, this boost can probably be attributed to the fact that the model uses many hand-tuned hyper-parameters that were selected to optimize the prediction of video quality on general-purpose content. CONTRIQUE, which is a deep learning model, was trained on a dataset of 2M images and performed the best among the frame-based single-model evaluators. CONTRIQUE's video-based counterpart, CONVIQT~\cite{madhusudana2022conviqt}, uses the CONTRIQUE backbone followed by a GRU to achieve incremental improvement. RAPIQUE, which is a hybrid model, performed best among all the evaluated VQA models by combining the handcrafted perceptual quality features with high-level semantic features generated by its deep learning module. 

\begin{figure*}[htbp]
    \centering
    {{\includegraphics[width=\textwidth]{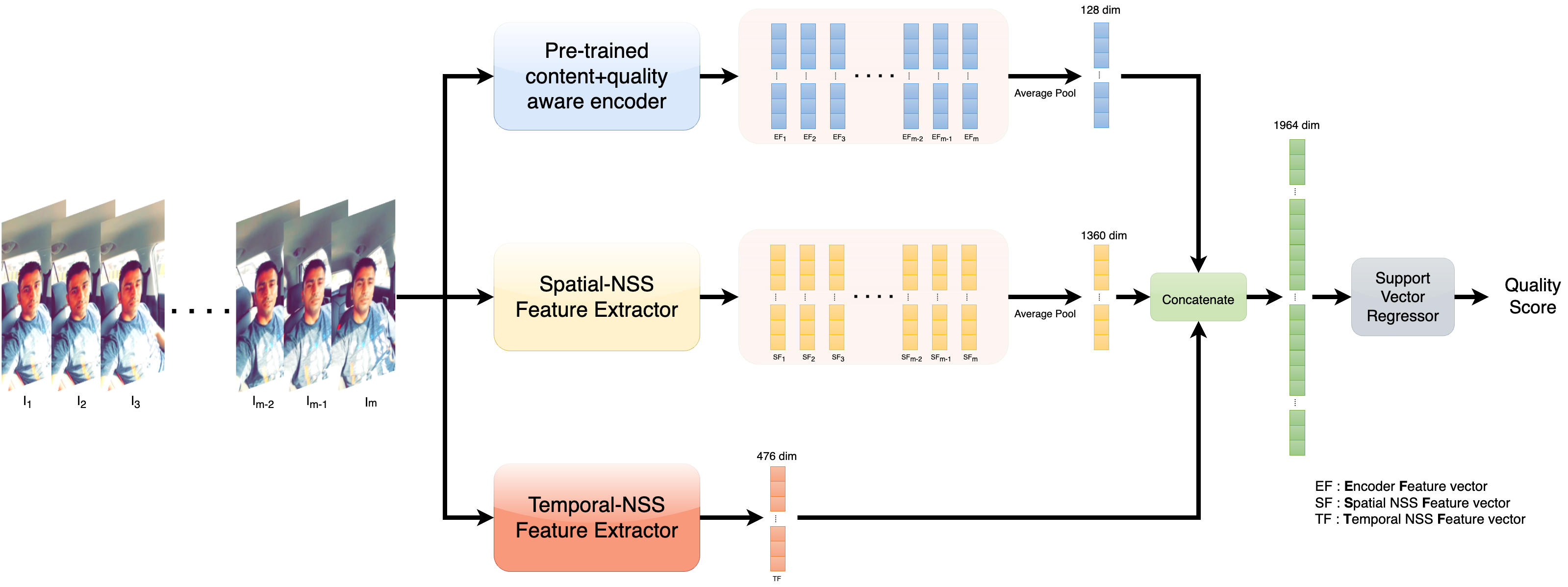}}}
    \captionsetup{justification=justified}
    \caption{MoEVA Evaluation Pipeline: Both the Pre-trained encoder and spatial NSS feature extractor operate at the frame level, thereby producing frame-level features. The temporal NSS feature extractor generates a single set of features using all the frames at once. The features computed over all frames are aggregated within their individual feature extractor blocks, wherever necessary, using simple average pooling, and then concatenated together before passing it through the regressor.}
    \label{fig:architectures_inference}
\end{figure*}

Although we noted that RAPIQUE performed the best among the compared models, the performances of VIDEVAL, CONTRIQUE, and other CNN-based methods such as PQR, Patch-VQ and Li \textit{et al.}~\cite{li2022blindly} were competitive. 

\section{Mixture-of-Expert based NR-VQA Model}
\label{sec:Moeva}
We have also developed a novel NR-VQA model called \textbf{MoEVA}, which is a \textbf{M}ixture-\textbf{o}f-\textbf{E}xperts-based \textbf{V}ideo-quality \textbf{A}ssessment algorithm. Partly inspired by RAPIQUE, MoEVA is a hybrid model that combines both distortion-aware spatial and temporal neurostatistical features with semantic-aware deep features. Unlike RAPIQUE, our approach to learning semantic information to augment perceptual quality prediction involves unsupervised training which can capture more general information than supervised training~\cite{grill2020bootstrap, caron2021emerging, henaff2021efficient}, helping the ultimate model perform better on various correlated tasks instead of specializing in one. 

A successful example of this is CONTRIQUE, which utilizes a contrastive training environment to train a model to learn to distinguish between different distortions, but with inferencing on quality judgments. CONTRIQUE uses an ancillary synthetic degraded image dataset generated by applying fixed distortions to a set of pristine images. Any two images impaired by the same type of distortion are categorized as the same, while any two images processed by different distortions are regarded as different, thus setting up the contrastive loss. CONTRIQUE improves the model's understanding of distortions by forcing it to learn them under a contrastive loss, but it hinders understanding of the image content by using the same loss. It forces the network to generate similar features for two images hvaing the same distortion settings even when their contents are different. To ameliorate this issue, we have developed a novel method of contrastive learning of distortions while better accounting for the impact of content. 

To conduct contrastive learning-based training, we require pairs of images that are labeled either as the same or as different. We also want to train the network on images at the same scale as when testing. Accordingly, we define our contrastive model at a patch level rather than on full frames. To be able to distinguish between patch labels and to create a protocol to label pairs accordingly, we lay out the following assumptions:

\begin{itemize}
    \item \textbf{A1}: The perceptual qualities of two neighboring patches are more likely to be similar than the perceptual qualities of two distant patches cropped from the same image. 
    \item[]
    \item \textbf{A2}: We also assume that the perceptual quality features computed from two patches taken from different images (contents) are different. 
    \item[]
    \item \textbf{A3}: Two versions of the same patch obtained by applying different types of distortion should have different perceptual quality features. Since the content within the two patches is exactly the same, any difference in the predicted perceptual quality should be reflected in the perceptual quality features generated by the VQA model.
\end{itemize}

These assumptions do not enforce any condition on quality scores, and different quality features can lead to similar scores.

Contrastive training is a self-supervised training strategy that learns representative features by understanding the relationships and differences between pairs of input images. We will notate paired samples as being either \textbf{+ve} examples or \textbf{-ve} examples. A \textbf{+ve} sample occurs when an input pair is labeled as similar/same, which encourages the model to generate similar features on the two input images. Conversely, a \textbf{-ve} sample occurs when the inputs in the pair are labeled as different, encouraging the model to generate different features. In the following subsections, we develop an augmented quality prediction scheme that relies on the generation of such labeled pairs under the aforementioned assumptions. 

\subsection{Training Dataset}
Since the 600 videos in the LIVE-ShareChat UGC-VQA database were curated from a larger publicly available set of 20,000 videos, it is reasonable to assume that the videos not included in the final dataset have similar features/characteristics as the ones that were in it. Thus we trained our model on the frames extracted from all 20,000 videos, excluding the 600 videos constituting the LIVE-ShareChat IUGC-VQA database. Since our assumptions A1-A3 are built on image specific properties, we extract all the frames from all the videos, and regard each frame as a distinct image in the training set. Since consecutive frames usually exhibit little change, instead of operating on every frame we only sample every $15^{th}$ frame and process the subsampled set.

\subsection{Content+Quality Aware Augmentation Scheme}
\begin{figure*}[htbp]
    \centering
    {{\includegraphics[width=\textwidth]{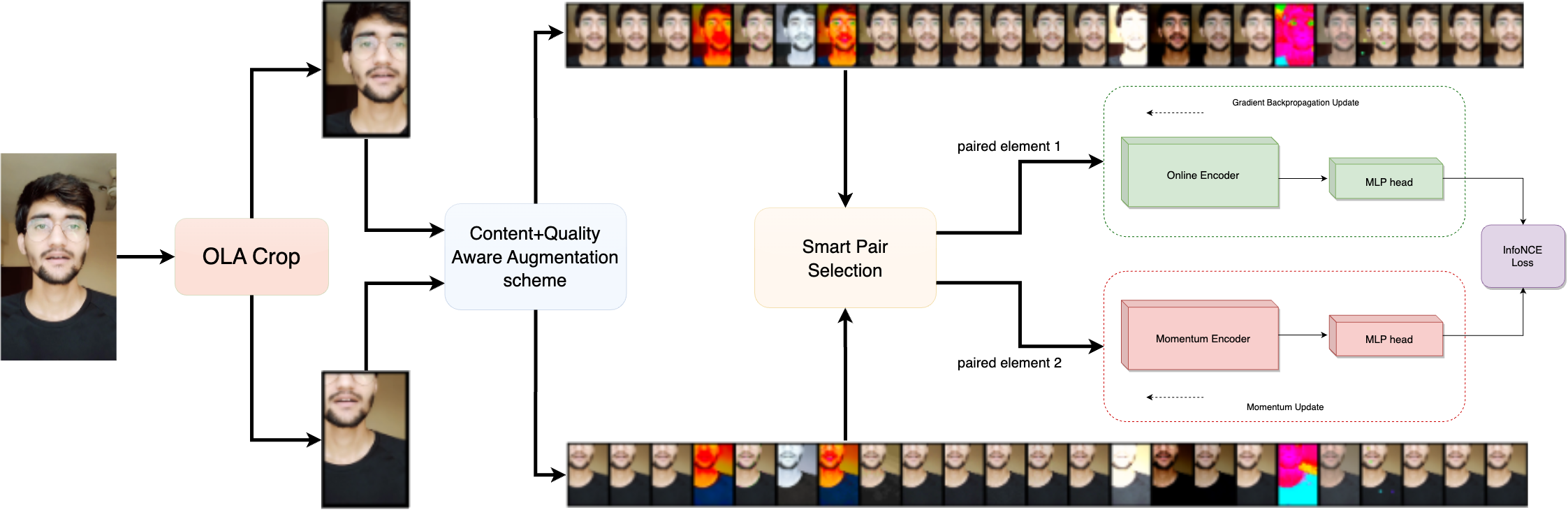}}}
    \captionsetup{justification=justified}
    \caption{MoEVA pre-training scheme: The overlapping area (OLA) based cropping mechanism selects two overlapping crops which are then augmented using the Content+Quality Aware Augmentation scheme. After pairing the augmented images under assumptions \textbf{A1-A3} in Section \ref{sec:Moeva}, pass the paired elements through the encoders, then the outputs of the encoders are passed through an MLP head which generates frame-level semantic features. Contrastive loss is applied on the representative features generated by both the Online and Momentum encoder.}
    \label{fig:architectures_training}
\end{figure*}
To enable the model to learn distinctions between different distortion settings, we deploy an augmentation bank that includes 25 distinct image-specific synthetic distortions, each applied at 5 different levels of severity. For any source frame $i^k$ from the training set, where $k \in {1, 2,...K,}$ and $K$ is the total number of images in the training data, a randomly chosen subset of the distortions available in the bank is applied to each image, resulting in a mini-batch of distorted images. We combine each source image $i^k$ with its generated augmentations to form a \textit{chunk}:

\begin{gather*}
    chunk^k = [i^k, i_1^k, i_2^k, ..., i_n^k],
\end{gather*}

\noindent where $i_j^k$ is the $j^{th}$ distorted version of $i^k$, and $n$ is the number of augmentations drawn from the bank. We then generate two random crops of $chunk^k$ denoted as $chunk^{k_c1}$ and $chunk^{k_c2}$, using an overlap area-based smart cropping mechanism. Specifically, crop locations are chosen such that the overlapping area (OLA) of the two crops falls within predetermined minimum and  maximum bounds. Ensure that the crop location is the same over all images in each chunk, but different between chunks, yielding:
\begin{gather*}
    \begin{split}
    chunk^{k_{c1}} &= [i^{k_{c1}},{i{_1}}^{k_{c1}}, {i{_2}}^{k_{c1}},..., {i{_{n}}}^{k_{c1}}]
    \end{split}
\end{gather*}
\noindent and
\begin{gather*}
    \begin{split}
    chunk^{k_{c2}} &= [{i}^{k_{c2}},{i{_1}}^{k_{c2}}, {i{_2}}^{k_{c2}},..., {i{_{n}}}^{k_{c2}}].
    \end{split}
\end{gather*}

After generating these augmentations, carefully pair and label them using previously stated assumptions, as follows:

\begin{gather*}
    \begin{split}
        [{i_{m}^{k_{c1}}}, {i_{m}^{k_{c2}}}] &\mapsto similar/same-quality\\
        [{i_{m}^{k_{c1}}},{i_{l}^{k_{c2}}}] &\mapsto different-quality\\
        [{i_{m}^{k_{c1}}}, {i_{l}^{k_{c1}}}] &\mapsto different-quality\\
        [{i_{m}^{k_{c1}}}, {i_{l}^{j_{c2}}}] &\mapsto different-quality.
    \end{split}
\end{gather*}

\subsection{Contrastive Pre-training}

Begin by defining two identical encoders 1) Online Encoder (O) and 2) Momentum Encoder (M). Both encoders are ResNet-50 backbones with an MLP head which generate the output embeddings used by the loss function. We split the pairs designed in the previous step, passing the first image in the pair through O and the other through M. Finally, to calculate the loss between the representations generated by O and M, we used the Noise-Contrastive Estimation \cite{gutmann2010noise} method in a manner that has proven effective in many other self-supervised contrastive learning paradigms \cite{henaff2020data,chen2020simple,he2020momentum,khosla2020supervised}. Specifically, we deploy the InfoNCE loss function as defined in \cite{oord2018representation}: 
\begin{equation}
    \!\!\mathcal{L}_{q, k^+, \{k^-\}} = -\log \frac{\exp(q \cdot k^+/\tau)}{\exp{(q \cdot k^+/\tau)}+\sum\limits_{k^-}^{} \exp(q \cdot k^-/\tau)}
    \label{eq:infonce}
\end{equation}

In Eq. \ref{eq:infonce} $q$ is the query image, $k^+$ is a positive sample (similar/same-quality), ${k^-}$ are the representations of the negative samples (different-quality, and $\tau$ is a temperature hyper-parameter. Optimization of this loss by updating the weights of O is conducted by backpropagation. The weights of M are updated using the weighted sum of the previous weights of M and the new weights of O. More formally denoting the parameters of O as $\theta_{O}$ and the parameters of M as $\theta_{M}$, update $\theta_{M}$ using:

\begin{equation}
    \theta_{M} \leftarrow m\theta_{M} + (1-m)\theta_{O},
\end{equation}

\noindent where $m\in[0,1)$ is the momentum coefficient. Once pre-training of the encoder has converged the ResNet-50 encoder weights of the Online encoder O can then be frozen and used for any downstream task associated with perceptual image quality.

\subsection{VQA Regression}

We compute representative features on each video by average pooling the frame-level features generated on each frame using the pre-trained encoder described in the previous subsection. These video features are then concatenated with the spatial and temporal NSS features, which are then collectively used to train an SVR head that learns to map the aggregate distortion and semantic-aware features to MOS. 

We also investigated the possibility of deploying a more sophisticated temporal pooling mechanism than simple average pooling the semantic features over frames. We implemented temporal pooling using a GRU, similar to CONVIQT~\cite{madhusudana2022conviqt}, but only obtained an insignificant performance boost as compared to simple average pooling. We believe this is because MoEVA already captures temporal quality variations via wavelet-based neurostatistical temporal features, which are of longer durations than frame-difference features used by many VQA models.

\begin{table}[]
\centering
\resizebox{\linewidth}{!}{%
\begin{tabular}{lcc|cc|cc}
\hline
\multirow{2}{*}{\textbf{Method}} & \multicolumn{2}{c|}{KoNViD-1k} & \multicolumn{2}{c|}{LIVE-VQC} & \multicolumn{2}{c}{YouTube-UGC} \\ \cline{2-7} 
          & \textbf{SRCC}   & \textbf{PLCC}   & \textbf{SRCC}   & \textbf{PLCC}   & \textbf{SRCC}   & \textbf{PLCC}   \\ \hline
NIQE      & 0.5392          & 0.5513          & 0.5930          & 0.6312          & 0.2499          & 0.2982          \\
BRISQUE   & 0.6493          & 0.6513          & 0.5936          & 0.6242          & 0.3932          & 0.4073          \\
VIDEVAL   & 0.7704          & 0.7709          & 0.7438          & 0.7476          & 0.7763          & 0.7715          \\
RAPIQUE   & 0.7884          & 0.8051          & 0.7413          & 0.7618          & 0.7473          & 0.7569          \\
Patch-VQ  & 0.7910          & 0.7860          & 0.8270          & 0.8370          & -               & -               \\
Li et al. & 0.8354          & 0.8339          & \textbf{0.8414} & \textbf{0.8394} & 0.8252          & 0.8178          \\
CONTRIQUE & 0.8440          & 0.8420          & 0.8150          & 0.8220          & 0.8250          & 0.8130          \\
CONVIQT   & \textbf{0.8510} & \textbf{0.8490} & 0.8080          & 0.8170          & \textbf{0.8320} & \textbf{0.8220} \\ \hline
MoEVA     & 0.8272          & 0.8314          & 0.8211          & 0.8225          & 0.6833          & 0.6845          \\ \hline
\end{tabular}%
}
\caption{Performance evaluation of MoEVA and prior NR-VQA algorithms on KoNViD-1k, LIVE-VQC, and YouTube-UGC}
\label{tab:moeva_vs_others}
\end{table}

\subsection{Experimental Results and Discussion}
We evaluated MoEVA in the same way as we evaluated the other algorithms. The specialized encoder performed substantially better than RAPIQUE's naive encoder, as shown in Table \ref{tab:performance_VQA}. For fair comparison, we built another model by combining CONTRIQUE with spatial and temporal neurostatistical features (S-NSS and T-NSS). Although this modified version of CONTRIQUE performed well, MoEVA performed significantly better, because of its enhanced semantic capacity.

We also compared the performance of MoEVA with prior VQA algorithms on other UGC-VQA databases in Table \ref{tab:performance_VQA}. While MoEVA performed competitively well on KoNViD-1k and LIVE-VQC, its performance on YouTube-UGC was relatively bad. It is also interesting to note that YouTube-UGC content and diversity features depicted in Fig. \ref{fig:freq_dist_and_tSNE} and \ref{fig:diversity_plots} are quite different from that of LIVE-ShareChat. Thus, MoEVA's poor performance on YouTube-UGC can be explained by MoEVA's deep learning component being trained on videos similar to that in LIVE-ShareChat.

\section{Conclusion}
Given the growing popularity of short-form videos on international social media platforms, we believe that the new LIVE-ShareChat IUGC-VQA Database is a valuable addition to the current collective of psychometric video quality datasets. Using the new resource, we benchmarked leading VQA algorithms and found significant room for improvement. Towards making progress on modeling, we also introduce a Mixture-of-Experts-based algorithm called MoEVA that combines spatial and temporal neurostatistical features with an unsupervised semantic deep learner to obtain complementary distortion-aware and content-aware perceptual quality features. The resulting MoEVA model significantly outperforms the compared SOTA algorithms on the LIVE-ShareChat IUGC-VQA database, but there remains room for improvement.

\appendices

\section*{Acknowledgment}

This research was sponsored by a grant from Mohalla Tech Pvt Ltd. The authors would like to thank ShareChat members: Sumandeep Banerjee and Mihir Goyal for providing their valuable feedbacks during discussions. The authors also thank the Texas Advanced Computing Center (TACC) at The University of Texas at Austin for providing compute resources that have contributed to the research results reported in this paper. URL:http://www.tacc.utexas.edu. 

\ifCLASSOPTIONcaptionsoff
  \newpage
\fi

\bibliographystyle{IEEEtran}

\bibliography{refs}

\end{document}